\documentclass[aps,superscriptaddress,amsmath,amssymb,floatfix,showpacs,notitlepage,nofootinbib,preprintnumbers]{revtex4-1}
\usepackage{graphicx,graphics,setspace,epsfig,color}
\usepackage[letterpaper, dvips,width=7.5in,height=8.5in,includemp=false]{geometry}
\usepackage[vcentermath]{}
\usepackage{epsf}
\usepackage{verbatim}
\usepackage{amscd}
\usepackage{amsmath}
\usepackage{graphicx}
\usepackage{dcolumn}
\usepackage{bm}
\usepackage{amssymb}
\usepackage{epsfig}
\usepackage{slashed}
\usepackage{lmodern} 
\usepackage[T1]{fontenc}

\newcommand{\be}{\begin{equation}}
\newcommand{\ee}{\end{equation}}
\newcommand{\bea}{\begin{eqnarray}}
\newcommand{\eea}{\end{eqnarray}}

\newcommand{\Ecm}{{E_{\rm cm}}}

\begin{document}

\preprint{INT-PUB-15-065}

\title[title]{Nucleon-nucleon bremsstrahlung of dark gauge bosons and revised supernova constraints.}
\author{Ermal Rrapaj}
 \email{ermalrrapaj@gmail.com}
\affiliation{Institute for Nuclear Theory, University of Washington, Seattle, WA}
\affiliation{Department of Physics, University of Washington, Seattle, WA}
\author{ Sanjay Reddy}
 \email{sareddy@uw.edu}
\affiliation{Institute for Nuclear Theory, University of Washington, Seattle, WA}
\affiliation{Department of Physics, University of Washington, Seattle, WA}

\begin{abstract}
We calculate the rate of production of hypothetical light vector bosons (LVBs) from nucleon-nucleon bremsstrahlung reactions in the soft radiation limit directly in terms of the measured nucleon-nucleon elastic cross sections.  We use these results and the  observation of neutrinos from supernova SN1987a to deduce constraints on the couplings of vector bosons with masses $\lesssim 200$ MeV to either electric charge (dark photons) or to baryon number. We establish for the first time strong constraints on LVB that couple only to baryon number, and revise earlier constraints on the dark photon. For the latter, we find that the excluded region of parameter space is diminished by about a factor of 10. 
\end{abstract}
\maketitle
\section{Introduction}
\label{section:intro}
The detection of about 20 neutrinos over about 10 seconds from supernova SN87a confirmed in broad-brush the paradigm for core-collapse supernova in which the neutrinos carry away the bulk of the gravitational binding energy $\simeq 3-5 \times 10^{53}$ ergs of the neutron star. The time scale associated with this intense neutrino emission is determined by neutrino diffusion in the hot and dense core of the newly born neutron star called the proto-neutron star\cite{Burrows:1986me}.  During this phase, the emission of other weakly interacting particles, were they to exist, could sap energy from the core and reduce the number and time scale over which neutrinos would be detectable. This allows one to extract useful constraints on the coupling of these hypothetical particles for masses up to about $200$ MeV from the neutrino signal observed from SN87a. Now widely referred to as the supernova cooling constraint \cite{Raffelt:1996wa}, it has provided stringent constraints on the properties of QCD axions \cite{Burrows:1988ah}, the size of large gravity-only extra-dimensions into which light Kaluza-Klein gravitons could be radiated \cite{Hanhart:2000ae,Hanhart:2001fx}, light supersymmetric particles such as neutralinos \cite{Dreiner:2003wh}, and more recently on the properties of dark photons \cite{Dent:2012mx,Dreiner:2013mua,Kazanas:2014mca}.  

Observations of galaxy rotation curves, the motion of galaxies in clusters, gravitational lensing, and the remarkable success of the $\Lambda$CDM model of the early universe (see Ref.~\cite{Olive:2003iq} for a pedagogic review), combined with the direct empirical evidence from the bullet cluster \cite{Clowe:2006eq} indicates the existence of dark matter (DM) which interacts with ordinary matter through gravitational interactions. This has spurred much recent research in particle physics and a plethora of DM models have been proposed that also naturally predict non-gravitational interactions. In a class of these models, DM is part of neutral hidden sector which interacts with standard model (SM) particles through the exchange of light vector bosons (LVBs) that couple to SM conserved currents \cite{Holdom:1986,Rajpoot:1989,Nelson:1989fx,Batell:2014yra}.  Here, DM is charged under a local $U(1)$ and from a phenomenological perspective, it is convenient to consider two possibilities. One in which the mediator couples to the SM electric charge $Q$, called the dark photon $\gamma_Q$ and is described by the spin-one field $A'_\mu$. The other in which the mediator couples only to baryon number, which is sometimes referred to as the leptophobic gauge boson $\gamma_B$ and is described by the field $B_\mu$. 

At low energy it suffices to consider minimal coupling of the LVBs to charge and baryon number described by the lagrangian
\begin{equation}
 \begin{split}
 \mathcal{L}\supset g_Q A'_\mu J_{\mu}^{\text{EM}} + g_B B_\mu  J_{\mu}^{\text{B}} -\frac{1}{2}m^2_{\gamma_Q} A'_\mu A'^{\mu}  -\frac{1}{2}m^2_{\gamma_B} B_\mu B^{\mu} \,, 
 \label{eq:lagrangian}
 \end{split}
\end{equation}
which also includes mass terms for the gauge bosons.  Of the two LVBs, the dark photon has been studied extensively and is usually discussed as arising from kinetic mixing of a dark sector gauge boson with the photon \cite{Jaeckel:2010ni}.  This mixing is described by the term $\epsilon_Q F'_{\mu\nu} F^{\mu\nu}$ in the low energy lagrangian where $F^{\mu\nu}$ and $F'_{\mu\nu}$ are the field tensors associated with the ordinary photon field and dark photon field, respectively. The Yukawa coupling in Eq.~\ref{eq:lagrangian}  $g_Q=\epsilon_Q e$ where $e=\sqrt{4\pi \alpha_{\rm em}}$ is the electric charge. To simplify notation, and for later convenience, we shall also introduce the parameter $\epsilon_B$ and write the Yukawa coupling of leptophobic gauge boson as $g_B=\epsilon_B e$. 

When the mass of the LVBs is less than or comparable to few times $T_{\rm SN}\simeq 30$ MeV, the temperature encountered in the supernova core, they can be produced copiously through nucleon-nucleon bremsstrahlung and electron-position pair-annihilation reactions. For both types of LVB, the bremsstrahlung production rate is expected to be the dominant contribution given the abundance of nucleons and the strong nature of nuclear interactions. In this article we calculate this production rate using the soft-radiation theorem and obtain a model independent estimate, related directly to the nucleon-nucleon elastic scattering data. A similar method was used in earlier work in \cite{Hanhart:2000ae} to estimate low energy neutrino and axion production and in \cite{Hanhart:2000er} to estimate the rate of production of Kaluza-Kelin gravitons and dilatons from nucleon-nucleon bremsstrahlung. Here we present for the first time a calculation of the rate of emission of the LVB $\gamma_B$ which couples to baryon number from nucleon-nucleon bremsstrahlung.  Our calculation of the bremsstrahlung production of dark photons predicts a rate that is about a  factor 10 smaller than that predicted in Ref.~\cite{Dent:2012mx}.  We trace this difference to an overly simplified treatment of the nucleon-nucleon interaction based on one-pion-exchange, and the use of the Born approximation for strong interactions.

In section \ref{section:soft} we review the well known result for soft bremsstrahlung radiation and outline the calculation for the emissivity of LVBs from the supernova core in this limit. We discuss the elastic neutron-neutron, proton-proton and neutron-proton cross-sections and use experimental data to compute the emissivities in section \ref{section:nn}. In section \ref{section:constraint} we derive constraints on $\epsilon_B$ and revise earlier constraints on $\epsilon_Q$. Here we also discuss sources of opacity for LVBs that can suppress cooling arising from inverse bremsstrahlung process, Compton scattering, and decay into electron--positron pairs. 

\section{Nucleon-nucleon bremsstrahlung in the soft limit}
\label{section:soft}
We begin by briefly reviewing nucleon-nucleon bremsstrahlung in the soft limit where the energy radiated is small compared to the energy associated with nucleon-nucleon interaction. It is well known that the amplitude for bremsstrahlung
production of particles can be related to the elastic scattering cross-section when expanded in powers of the energy $\omega$, 
carried away by the radiated particles\cite{bjorken:1964}. The amplitude for a generic bremsstrahlung process $XY \rightarrow XY \gamma$ can be written as 
\be
{\cal M}_{XY\rightarrow XY\gamma}= \frac{A(E_{\rm cm})}{\omega}+B(E_{\rm cm})+ {\cal O}(\omega) \,,  
\ee 
where $A(E_{\rm cm})$ and $B(E_{\rm cm})$ are related directly to the elastic $XY \rightarrow XY$ cross-section without radiation in the final state.  This result, called Low's soft-photon theorem for bremsstrahlung was first derived  by F. E. Low \cite{Low:1958} and has been used to study neutron-proton and proton-proton bremsstrahlung reactions since the pioneering work of \cite{Nyman:1968,Heller:1968}. Calculations of the bremsstrahlung rate in which only terms arising from on-shell elastic amplitudes $A(E_{\rm cm})$ and $B(E_{\rm cm})$ is generally referred to as the soft-photon approximation or the soft radiation approximation (SRA).  

\begin{figure}[htbp] 
   \centering
   \includegraphics[width=4in]{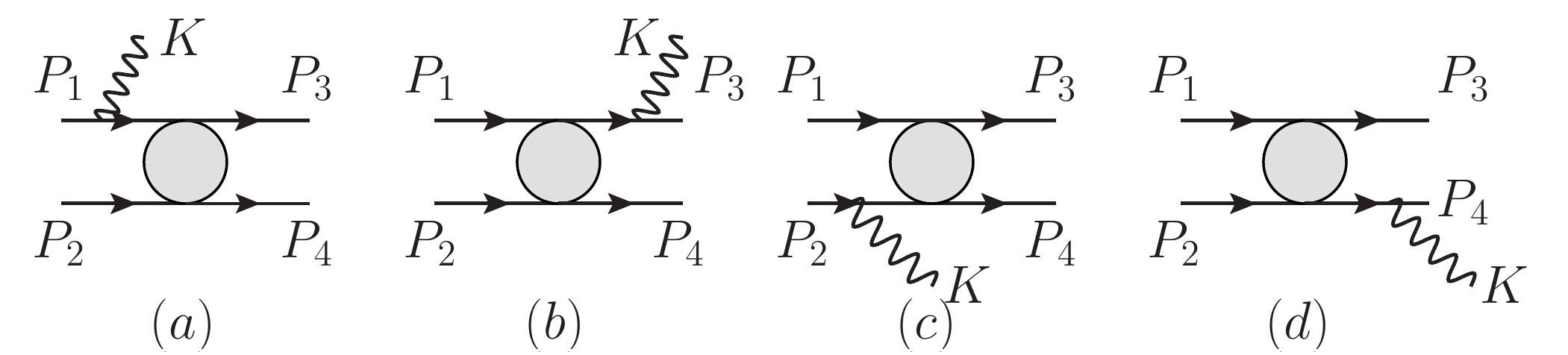} 
   \caption{Diagrams in which radiation denoted by the wavy-line attaches to the external nucleon legs (solid lines)  dominates 
   in the low energy limit. The grey blob represents the anti-symmetrized nucleon-nucleon potential and contains both the direct and exchange contributions. }
   \label{fig:external}
\end{figure}

The Feynman diagrams that contribute in the SRA  are shown in Fig.~\ref{fig:external}.  Here nucleons are represented by solid lines, the LVB as the wavy-photon lines and the shaded circle represents the nucleon-nucleon interaction which contains both the long-distance component arising from pion-exchanges and all of the effects of the short-distance components that contribute to nucleon-nucleon scattering. 
The amplitude for the reaction $pp\rightarrow pp \gamma$ is obtained by summing diagrams $(a),(b),(c)$ and $(d)$, while for the reaction $np\rightarrow np \gamma$ only two of these diagrams contribute in which the photon couples only to the proton at leading order in this expansion.  The four momenta of the initial state nucleons is denoted $P_1$ and $P_2$, and by $P_3$ and $P_4$ in the final state. $K=(\omega,\vec{k})$ is the four momentum of the radiated quanta and $\epsilon_{\mu}$ is its polarization. These diagrams dominate at small $\omega$ because the intermediate nucleon is close to being on-shell and makes a contribution to the bremsstrahlung amplitude at order $\omega^{-1}$.  In this limit, when the energy radiated is small compared to $E_{\rm cm}$ of the nucleon pair, the unpolarized differential cross-sections for bremsstrahlung radiation of LVBs are given by 
\bea
d\sigma_{pp \rightarrow pp \gamma_i} &=&-4 \pi \alpha_{\rm em}\epsilon_i^2~ \frac{d^3k}{2\omega}(\epsilon^{\mu} J^{(4)}_\mu)^2~ d\sigma_{pp\rightarrow pp} \,, 
\label{eq:softa} \\
d\sigma_{np \rightarrow pp \gamma_Q} &=&-4 \pi \alpha_{\rm em}\epsilon_Q^2~ \frac{d^3k}{2\omega}(\epsilon^{\mu} J^{(2)}_\mu)^2~ d\sigma_{np\rightarrow np} \,, 
\label{eq:softb} \\
d\sigma_{np \rightarrow np \gamma_B} &=&-4 \pi \alpha_{\rm em}\epsilon_B^2 ~\frac{d^3k}{2\omega}(\epsilon^{\mu} J^{(4)}_\mu)^2~ d\sigma_{np\rightarrow np} \,,
\label{eq:softc}
\eea   
where 
\bea  
J^{(2)}_\mu&=&\left(\frac{P_1}{P_1\cdot K} - \frac{P_3}{P_3\cdot K} \right)_\mu \,,\\
J^{(4)}_\mu&=&\left(\frac{P_1}{P_1\cdot K}+\frac{P_2}{P_2\cdot K} - \frac{P_3}{P_3\cdot K} - \frac{P_4}{P_4\cdot K}\right)_\mu \,,
\eea
are the currents associated with dipole and quadrupole radiation, respectively \cite{bjorken:1964,Nyman:1968}. The unpolarized elastic differential cross-sections for $pp$ and $np$ and given by $d\sigma_{pp\rightarrow pp}$ and $d\sigma_{np\rightarrow np}$, respectively.  These results are valid to leading order (LO) in an expansion in powers of $\chi=\omega/E_{\rm cm}$ where $\Ecm=(\vec{p}_1-\vec{p}_2)^2/4M$ is the non-relativistic center of mass (cm) energy.  When it is appropriate to only retain terms at order $\chi^{-2} $ the elastic cross-section $d\sigma$ is calculated at the $E_{\rm cm}$ and is determined by the incoming nucleon energies.  Next-to-leading order corrections at order $\chi^{-1} $  and $\chi^{0}$ arise and are proportional to the $d\sigma/d\Ecm$ and can be come important when $\Ecm \lesssim 10$ MeV where $d\sigma$ varies rapidly. However, for ambient conditions in the supernova core $\Ecm \approx 100$ MeV and for these energies $d\log{\sigma}/d\log{\Ecm} \ll 1$ and these corrections can be expected to be small.  

\begin{figure}[htbp] 
   \centering
   \includegraphics[width=4in]{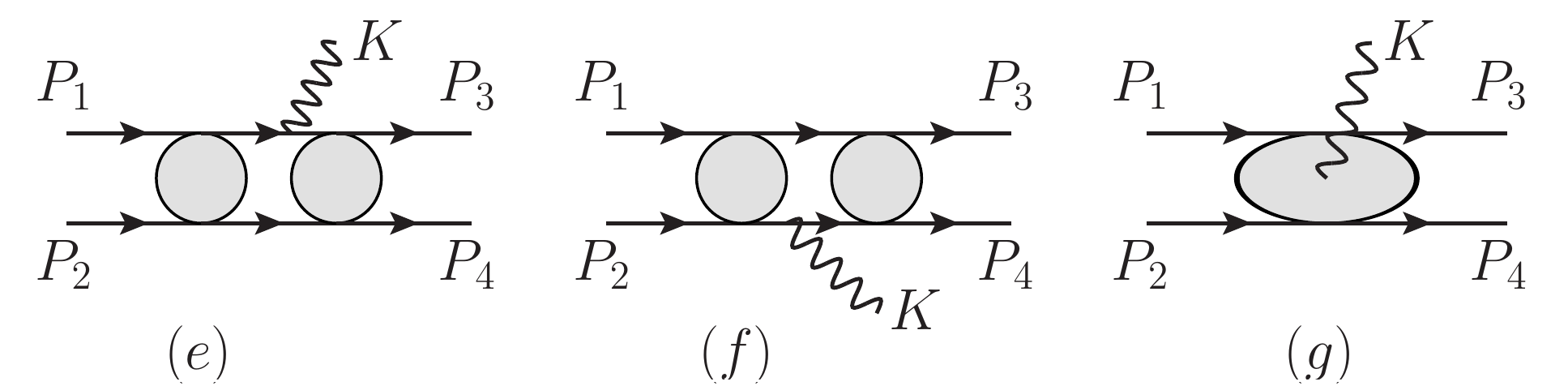} 
   \caption{Neglected diagrams $(e)$ and $(f)$ in which radiation attaches to internal nucleon lines, and $(g)$ in which it couples to short-distance two-body currents represented by the grey blob. }
   \label{fig:internal}
\end{figure}

Diagrams shown in Fig.~\ref{fig:internal} contribute to bremsstrahlung radiation at order $\chi^1$ in the low energy expansion.   Here, the separation between the contributions from diagrams labelled $(e)$ and $(f)$, and the two-body current shown in the diagram labelled $(g)$ is model and scale dependent and it is inconsistent to selectively include any subset of these contributions. We also note, once again, that the grey blobs should include both the pion exchanges and short-distance contributions and latter being especially important.  Comparisons between model calculations which include order $\chi$ contributions with those obtained in the SRA,  and nucleon-nucleon bremsstrahlung data find that the SRA provides as good a description of the data as do the potential models with their prescribed 2-body currents\cite{Huisman:1999a}. For this reasons we will neglect the contributions from the diagrams in Fig.~\ref{fig:internal} and use Eqns.~\ref{eq:softa}, \ref{eq:softb} \& \ref{eq:softc} to calculate the emission rates. A comparison between the photon bremsstrahlung data measured in the laboratory,  and predictions of the rate in SRA provides an estimate of the associated error.  For collisions with $\Ecm \approx 100$ MeV these comparisons show that the SRA provides a good description of the data for $\omega << \Ecm$, and for $\omega \simeq \Ecm$ underestimates the cross-sections by about a factor of about 2  \cite{Huisman:1999a,Huisman:1999b,Safkan:2007iy}.  For these reasons we consider the leading order SRA better suited to calculate emission and scattering rates of LVB rather than models which include corrections arising from a sub-class diagrams in Fig.~\ref{fig:internal} in perturbation theory.

The emissivity, which is the rate of emission of energy in LVBs per unit volume, can be calculated in the SRA using Eqns.~\ref{eq:softa}, \ref{eq:softb}, and \ref{eq:softc}. For the process $np\rightarrow np\gamma_Q$ and $np\rightarrow np\gamma_B$ they are given by   
\bea 
\dot{\epsilon}_{np\rightarrow np \gamma_Q} &=& -4\pi \alpha_{\rm em} \epsilon_Q^2 \int \frac{d^3k}{2\omega (2\pi)^3} \omega  \int \frac{d^3p_1 f_n(E_1)}{(2E_1)(2\pi)^3} \int \frac{d^3p_2f_p(E_2)}{(2E_2)(2\pi)^3} \int d\Pi (\epsilon^{\mu} J^{(2)}_\mu)^2 ~32 \pi E^2_{\rm cm}v_{rel}~\frac{d\sigma_{\rm np}(E_{\rm cm},\theta)}{d\theta_{\rm cm}} \,,
\label{eq:emis_Q}
\\
\dot{\epsilon}_{np\rightarrow np \gamma_B} &=& -4\pi \alpha_{\rm em} \epsilon_B^2 \int \frac{d^3k}{2\omega (2\pi)^3} \omega \int \frac{d^3p_1f_n(E_1)}{(2E_1)(2\pi)^3}  \int \frac{d^3p_2f_p(E_2) }{(2E_2)(2\pi)^3} \int d\Pi (\epsilon^{\mu} J^{(4)}_\mu)^2 ~32 \pi E^2_{\rm cm}~v_{rel}~\frac{d\sigma_{\rm np}(E_{\rm cm},\theta)}{d\theta_{\rm cm}} \,,
\label{eq:emis_B}
\eea
where 
\be 
d\Pi = (2\pi)^4~\delta^4(P_1+P_2-P_3-P_4-k) (1-f_n(E_3))(1-f_p(E_4))\frac{d^3p_3}{2E_3(2\pi)^3}\frac{d^3p_4}{2E_4(2\pi)^3}\,,
\label{eq:phasespace}
\ee
is the final state phase space of the nucleons, $d\sigma_{\rm np}/d\theta$ is the differential elastic $np$ scattering cross-section, $v_{rel}= |\vec{p}_1-\vec{p}_2|/M$ is the relative speed, and $\theta_{\rm cm}$ is the scattering  angle. $f_i(E)=1/(1+\exp{((E-\mu_i)/T)})$ is the Fermi distributions functions for neutrons and protons.  Eq.~\ref{eq:phasespace} includes Pauli blocking factors for the final state nucleons and is important under degenerate conditions. However, in the supernova core, matter is partially degenerate with $\mu_{(n/p)}/T\simeq 1$ and under these conditions the suppression due to Pauli blocking is small. The emission rates due to the reactions $nn\rightarrow nn\gamma_B$ and $pp\rightarrow pp\gamma_B$   are obtained by replacing $d\sigma_{\rm np}$ in Eq.~\ref{eq:emis_B} by $d\sigma_{\rm nn}$ and $d\sigma_{\rm pp}$, respectively and introduce the relevant distribution functions. Similarly to obtain the contribution for the reaction $pp\rightarrow pp\gamma_Q$ we replace $d\sigma_{\rm np}$ in Eq.~\ref{eq:emis_Q} by $d\sigma_{\rm pp}$ and $f_n$ by $f_p$.  In section \ref{section:nn} we discuss our calculations of the elastic nucleon-nucleon cross-sections and find that since  $d\sigma_{\rm np}$ is larger at the energies of interest and because $\gamma_Q$ radiation occurs at dipole order in the $np$ reaction, the quadrupole order contribution from the $pp\rightarrow pp\gamma_Q$  reaction is small. 

Despite the high density and temperature in the supernova core the typical nucleon velocity $v \lesssim 1/3$ and it is useful to expand in powers of $v$ as this leads to simplifications that allows us to do the phase space integrals needed to calculate the emissivities. We find that under non-degenerate conditions the emissivities are given by the following compact formulae 
\bea
  \dot{\epsilon}_{np\rightarrow np\gamma_Q}&=&\frac{\alpha_{\rm em}\epsilon_Q^2}{{\pi}^{3/2}} \frac{n_n n_p}{ (MT)^{3/2}}\int_{m_{\gamma_Q}}^{\infty} d\Ecm ~e^{-\frac{\Ecm}{T}}~ E^3_{\text{cm}}~ \mathcal{I}^{(2)}(\frac{m_{\gamma_Q}}{\Ecm}) ~ \sigma^{(2)}_{np}(\Ecm) 
   \label{eq:rate_npQ}
  \\
   \dot{\epsilon}_{pp\rightarrow pp\gamma_Q}&=&\frac{\alpha_{\rm em}\epsilon_Q^2}{{\pi}^{3/2}} \frac{n_p n_p}{ (MT)^{3/2}}\int_{m_{\gamma_Q}}^{\infty}dE_{\text{cm}} ~e^{-\frac{\Ecm}{T}}~\frac{E_{\text{cm}}^4 }{M} ~\mathcal{I}^{(4)}(\frac{m_{\gamma_Q}}{\Ecm}) ~ \sigma^{(4)}_{pp}(\Ecm) 
  \label{eq:rate_npB} \\
  \dot{\epsilon}_{ij\rightarrow ij\gamma_B}&=&\frac{\alpha_{\rm em}\epsilon_B^2}{{\pi}^{3/2}} \frac{n_i n_j}{ (MT)^{3/2}}\int_{m_{\gamma_B}}^{\infty}dE_{\text{cm}} ~e^{-\frac{\Ecm}{T}} ~\frac{E_{\text{cm}}^4 }{M} ~\mathcal{I}^{(4)}(\frac{m_{\gamma_B}}{\Ecm})~ \sigma^{(4)}_{ij}(\Ecm) 
  \label{eq:rate_npB} 
\eea
where
\bea
  \mathcal{I}^{(2)}(x)&=&\frac{4}{3} \left( \sqrt{1-x^2}(1-\frac{x^2}{4})- \frac{3x}{4} \arctan{\left(\frac{\sqrt{1-x^2}}{x} \right)} \right) \,,  \\  
  \mathcal{I}^{(4)}(x)&=&\frac{8}{5} \left( \sqrt{1-x^2}(1+\frac{x^2}{12}+\frac{x^4}{6})-\frac{5x}{4} \arctan{\left(\frac{\sqrt{1-x^2}}{x} \right)} \right)\,,
  \label{eq:currents}
\eea
and
\bea
  \sigma^{(2)}_{ij}&=&\int d\cos \theta_{\text{cm}} \frac{d\sigma_{n_in_j\rightarrow n_in_j}}{d\theta_{\text{cm}}} (1-\cos \theta_{\text{cm}}) \,, 
  \label{eq:sig_2}
  \\
  \sigma^{(4)}_{ij}&=&\int d\cos \theta_{\text{cm}} \frac{d\sigma_{n_in_j\rightarrow n_in_j}}{d\theta_{\text{cm}}} (1-\cos^2 \theta_{\text{cm}})\,. 
\label{eq:sig_4}  
\eea
The derivation of these results is discussed in Appendix \ref{appendix:emissivity}. Albeit cumbersome, numerical calculations of the emissivity including relativistic dispersion relations for the nucleons and corrections due to matter degeneracy can be performed directly using Eqs.~\ref{eq:emis_Q} and  \ref{eq:emis_B}.  At $T=30$ MeV and nucleon number density  $n\simeq n_0=0.16$ fm$^{-3}$ we have estimated these corrections to be small $\simeq 30\%$ compared to order $\chi$ corrections neglected in the SRA, which could be about factor of 2 as mentioned earlier.
\section{Elastic cross-sections and LVB emissivity}
\label{section:nn}
\begin{figure}[ht]
 \begin{center}
  \includegraphics[scale=0.5]{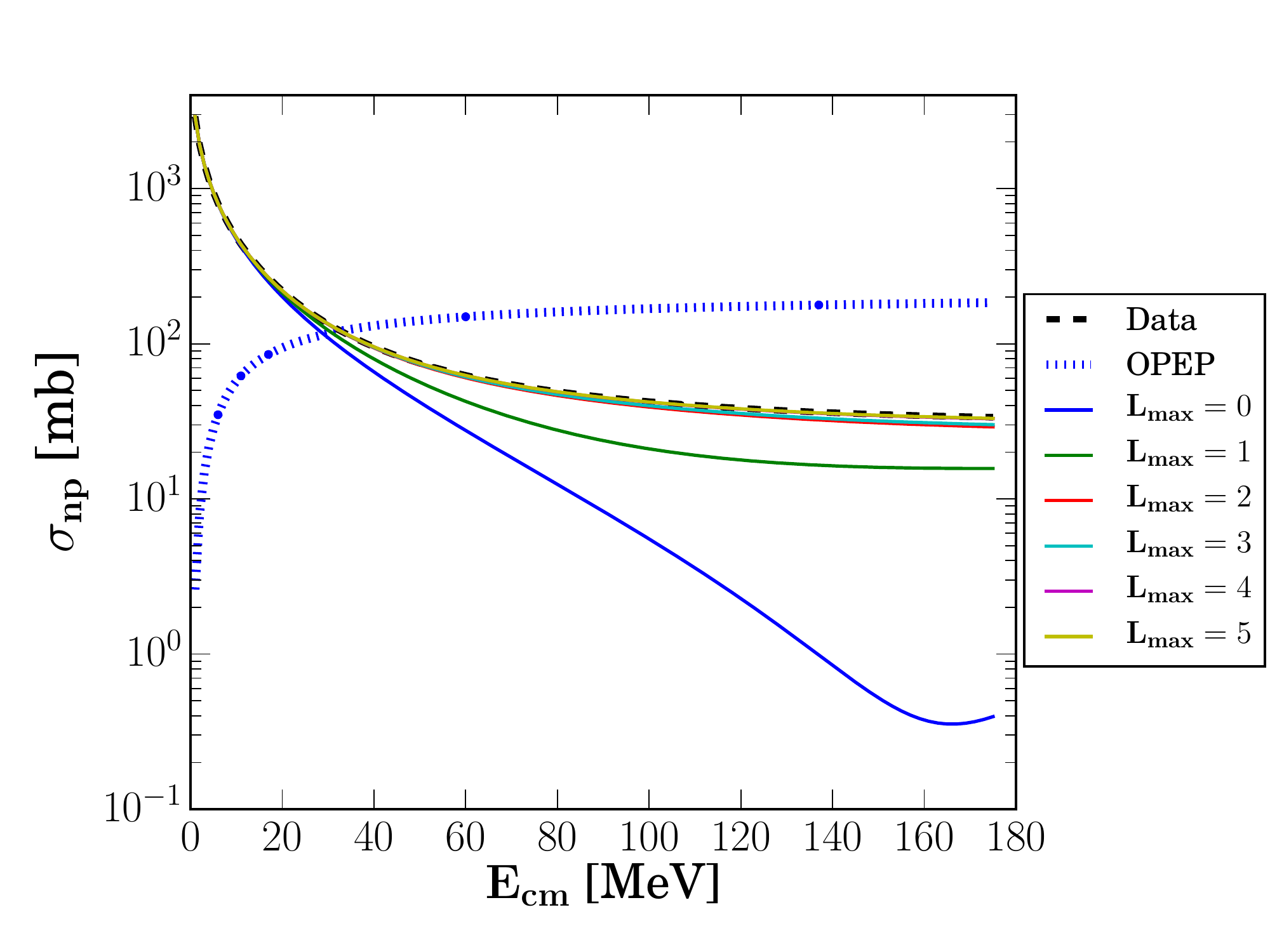}
  \end{center}
  \caption{The np scattering cross section  reconstructed from the phase shifts  is compared with data and the predictions of the OPEP. The change in the total cross-section as more partial waves are included is shown and is accordance with the expectation about its rapid convergence. In contrast, the Born cross-sections in the OPEP fail to reproduce both the qualitative and quantitative features seen in the data.}
  \label{fig:sigma_np_compare}   
 \end{figure}

Any realistic nucleon-nucleon potential constructed to reproduce the nucleon-nucleon phase shifts can be used to calculate the elastic differential cross-sections appearing in Eqs.~\ref{eq:sig_2} and \ref{eq:sig_4}.  
It is, however, simpler to obtain these cross-sections directly from the measured phase shifts and this calculation is outlined in Appendix~\ref{section:appendix}.
The differential cross section is expanded in the spherical wave basis with definite orbital angular momentum $L$ and the angular dependence is given by associated Legendre polynomials $P_l^{m}(\cos \theta_{\text{cm}})$ and the energy dependence is encoded in the phase shifts \cite{PWA}.
This expansion converges rather rapidly as can be seen from Fig.~\ref{fig:sigma_np_compare} where we present our calculation of the total $n-p$ cross-section including individual contributions from phase shifts with angular momentum from $L =0$ to $L=5$. With the inclusion of phase shifts $L=0,1,\&,2$ one finds good agreement between theory and the high quality data shown by the dashed curve. To better resolve the angular dependancies needed to determine  $\sigma^{(2)}_{ij}$ and $\sigma^{(4)}_{ij}$ we retain terms up to $L=5$.

In Fig.~\ref{fig:sigma_np_compare}  we also show the $np$ cross-section calculated in the Born approximation using the one pion exchange potential (OPEP).  
A comparison reveals large differences in the magnitude and energy dependence of the cross-section and implies that earlier work in Ref.~\cite{Dent:2012mx} where bremsstrahlung was calculated using the one pion exchange potential will also be similarly discrepant. We can deduce that at small values of the $\Ecm$ one pion exchange model grossly underestimates the scattering rate, while for  $\Ecm > 50 $ MeV it overestimates it by about a factor of $6$. We find a similar trend for the $nn$ and $pp$ cross-sections. In the supernova where $T\simeq 30$ MeV the relevant $\Ecm \simeq100$ MeV, and we can anticipate that calculations based on the OPEP will overestimates the bremsstrahlung rate by a similar factor. As we shall see shortly this is borne out by the comparison between our results for the dark photon production with those presented in \cite{Dent:2012mx}. 
\begin{center}
 \begin{figure}[ht]
  \includegraphics[scale=0.45]{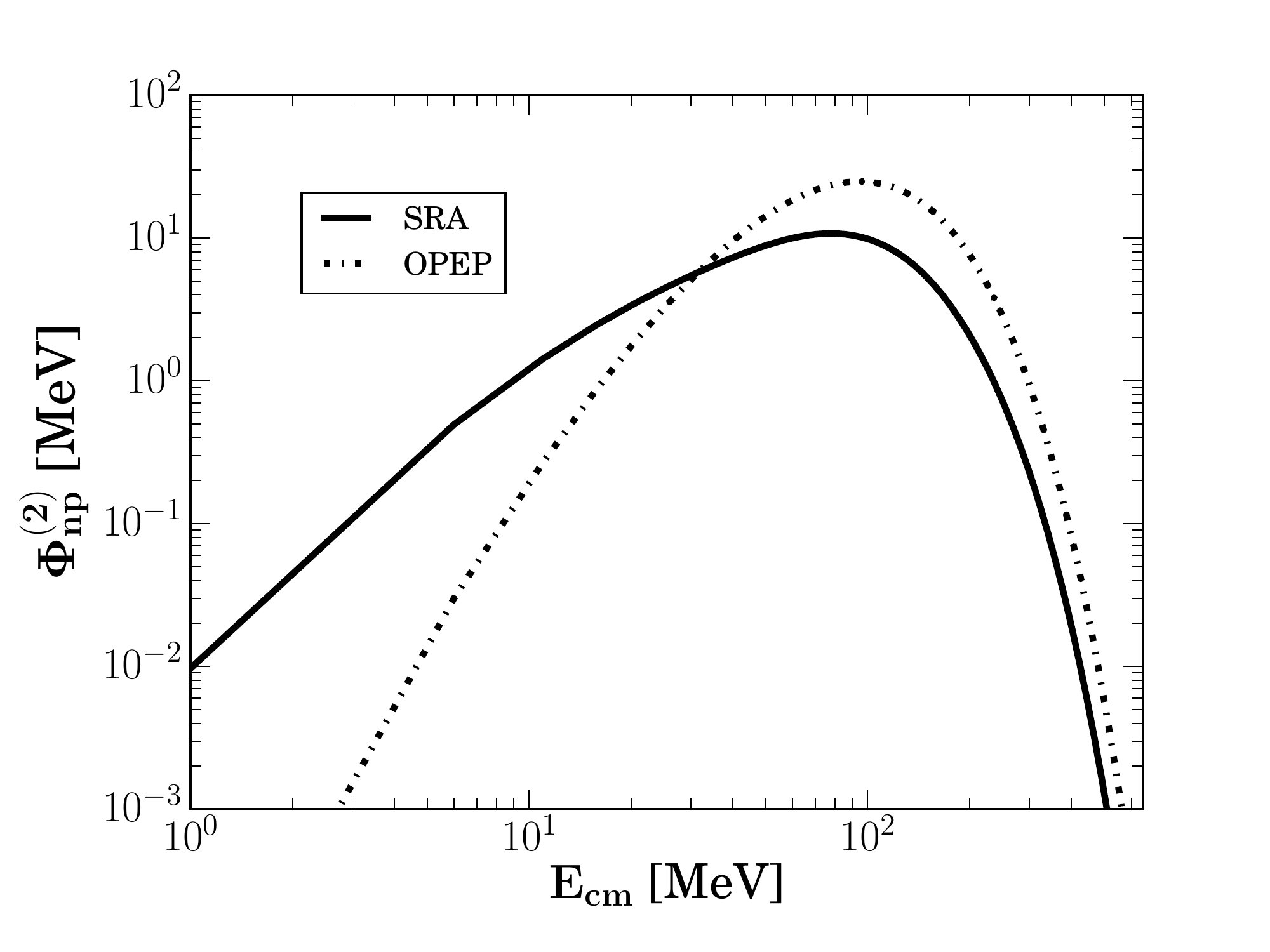}
  \includegraphics[scale=0.45]{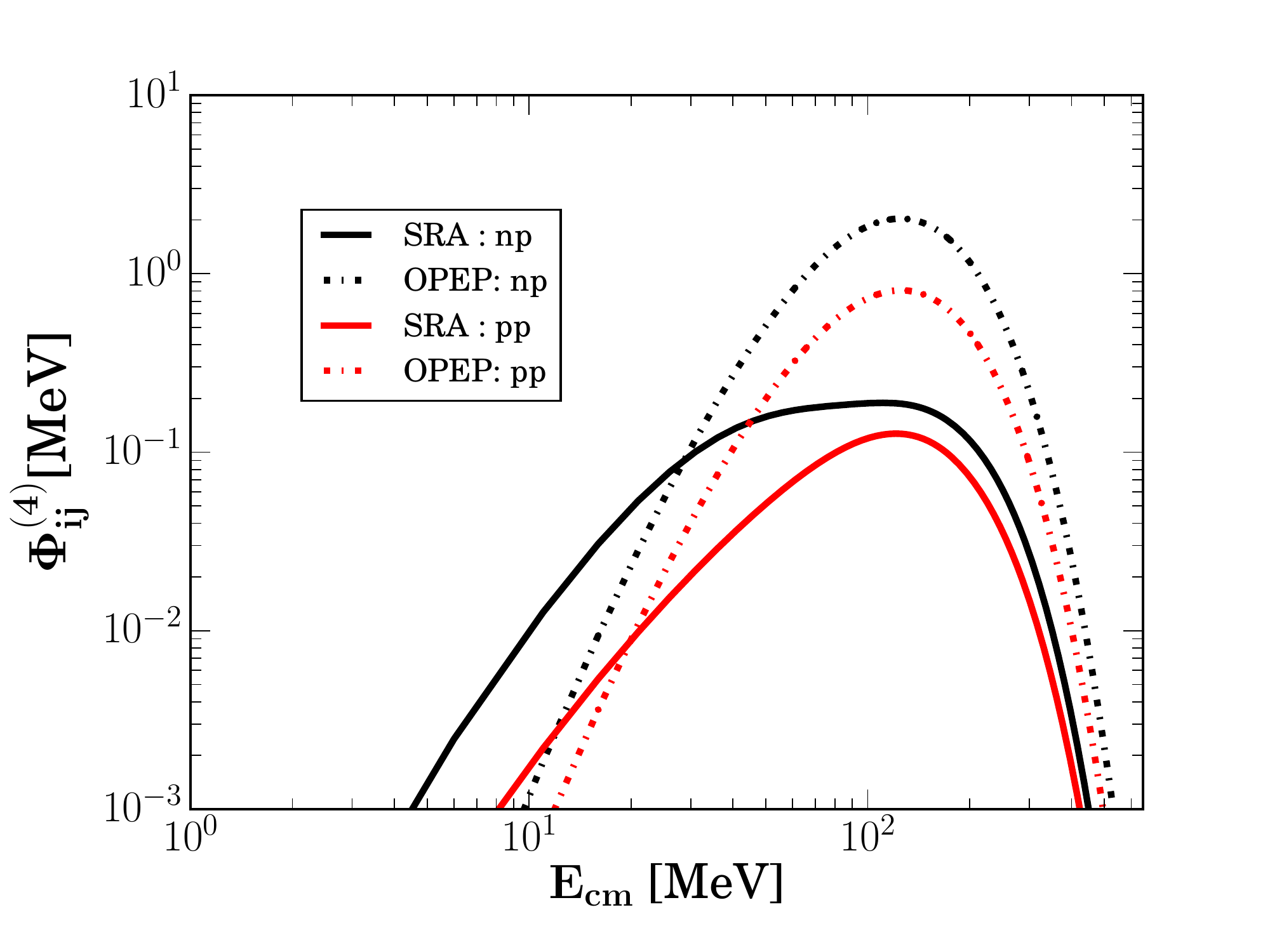}
  \caption{Dipole (left) and quadrupole (right) contributions to the emissivity integral defined in Eqs.~\ref{eq:phi_2} \& \ref{eq:phi_4}, respectively. The solid lines were obtained using experimentally measured differential cross sections and the dotted lines are obtained from the OPEP model. Fiducial values $T = 30$ MeV and $m=0$ are used in these plots.}
  \label{fig:phasespacesigma}
  \end{figure}
\end{center}

In the left panel of Fig.~\ref{fig:phasespacesigma} we plot the function
\be 
\Phi_{np}^{(2)}(\Ecm)=\exp{\left(-\frac{\Ecm}{T}\right)}~ E^3_{\text{cm}}~ \mathcal{I}^{(2)}(0) ~ \sigma^{(2)}_{np}(\Ecm) 
\label{eq:phi_2}
\ee
which appears as the integrand on the RHS of  Eq.\ref{eq:rate_npQ}, and in the right panel we show the function  
\be 
\Phi_{ij}^{(4)}(\Ecm)=\exp{\left(-\frac{\Ecm}{T}\right)}~ \frac{E^4_{\text{cm}}}{M}~ \mathcal{I}^{(4)}(0) ~ \sigma^{(4)}_{ij}(\Ecm) 
\label{eq:phi_4}
\ee
which appears as the integrand on the RHS of Eq.\ref{eq:rate_npB} with $m=0$.  These plots show the distribution of center of mass energies of nucleons in the initial state that contribute to the bremsstrahlung process when the mass of the LVB is negligible.  The $np\rightarrow np \gamma_Q$ process shown in the left panel is stronger because this occurs at dipole order, while the processes that occur at quadrupole order $pp\rightarrow pp \gamma_Q$, $np\rightarrow np \gamma_B$, $nn \rightarrow nn \gamma_B$ and $pp \rightarrow pp \gamma_B$  are shown in the right panel are suppressed by $\Ecm/M\propto v^2$ where $v$  is nucleon velocity in the initial state.  The emissivity is proportional to the area under these curves and difference between the curves obtained in the SRA and the OPEP is striking and the trends follow from the comparison between the cross sections seen in Fig.~\ref{fig:sigma_np_compare}. For soft dipole radiation these curves suggest that OPEP would overestimate the rate by about a factor $2$, while for quadrupole radiation it would overestimate the rate by about a factor $10$. In these plots $T=T_{SN}=30$ MeV and under these conditions we see that bremsstrahlung production of dark photons peaks at  $\Ecm\simeq 100$ MeV and production of leptophobic LVB peaks at $\Ecm\simeq 150$ MeV. The spectrum of LVBs emitted will be approximately thermal with $\omega \approx T-3T$ suggesting that expansion parameter for the SRA, $\chi \simeq 1/5-1$. 

\section{New and Revised Constraints}
\label{section:constraint}
In earlier work Raffelt found empirically that when the energy loss rate per gram due to the radiation of free streaming particles in the supernova core at a fiducial density  $\rho=3\times 10^{14} ~\text{g}/ \text{cm}^{3}$ and temperature $T = 30$ MeV  exceeds 
\be 
\dot{E}_{\rm Raffelt} = 10^{19}~ \frac{\text{erg}}{\text{g}~ \text{s}}
\ee
the duration of the SN neutrino burst is approximately reduced by half  \cite{Raffelt:1996wa}. Detailed simulations of neutrino transport in the protoneutron star and its predictions for the neutrino events in  Kamioka and IMB which were the neutrino detectors at the time of SN87a validate Raffelt's approximate local criterion \cite{Raffelt:1996wa,Hanhart:2001fx} and in what follows we shall employ it to constrain $\epsilon_Q$ and $\epsilon_B$.  We note that Raffelt's criterion approximately corresponds to limiting the energy loss due to LVBs to total luminosity of $L < \dot{E}_{\rm Raffelt}\times M_{\rm core} \simeq 2\times 10^{52}~(M_{\rm core}/M_\odot)$ ergs/s. 

First, we determine the SN87a constraints on $\gamma_B$, which is the leptophobic LVB that couples to baryon number. The total energy loss rate per gram due to $\gamma_B$ radiation is 
\be
\dot{E}_B(\rho,T,Y_p) = (\dot{\epsilon}_{np\rightarrow np\gamma_B}+\dot{\epsilon}_{nn\rightarrow nn\gamma_B}+ \dot{\epsilon}_{pp\rightarrow pp\gamma_B})/\rho \,,
\ee
where $\rho$ is the matter mass density, $T$ is the temperature and $Y_p=n_p/(n_n+n_p)$ is the fraction of protons. As already noted we choose $\rho=3\times 10^{14} ~\text{g}/ \text{cm}^{3}$, $T = T_{\rm SN}=30$ MeV and we set the proton fraction $Y_p=0.3$ to reflect typical conditions encountered in proto-neutron star simulations\cite{Keil:1994sm,Pons:1998mm}. 

\begin{center}
\begin{figure}[h]
 \includegraphics[scale=0.6]{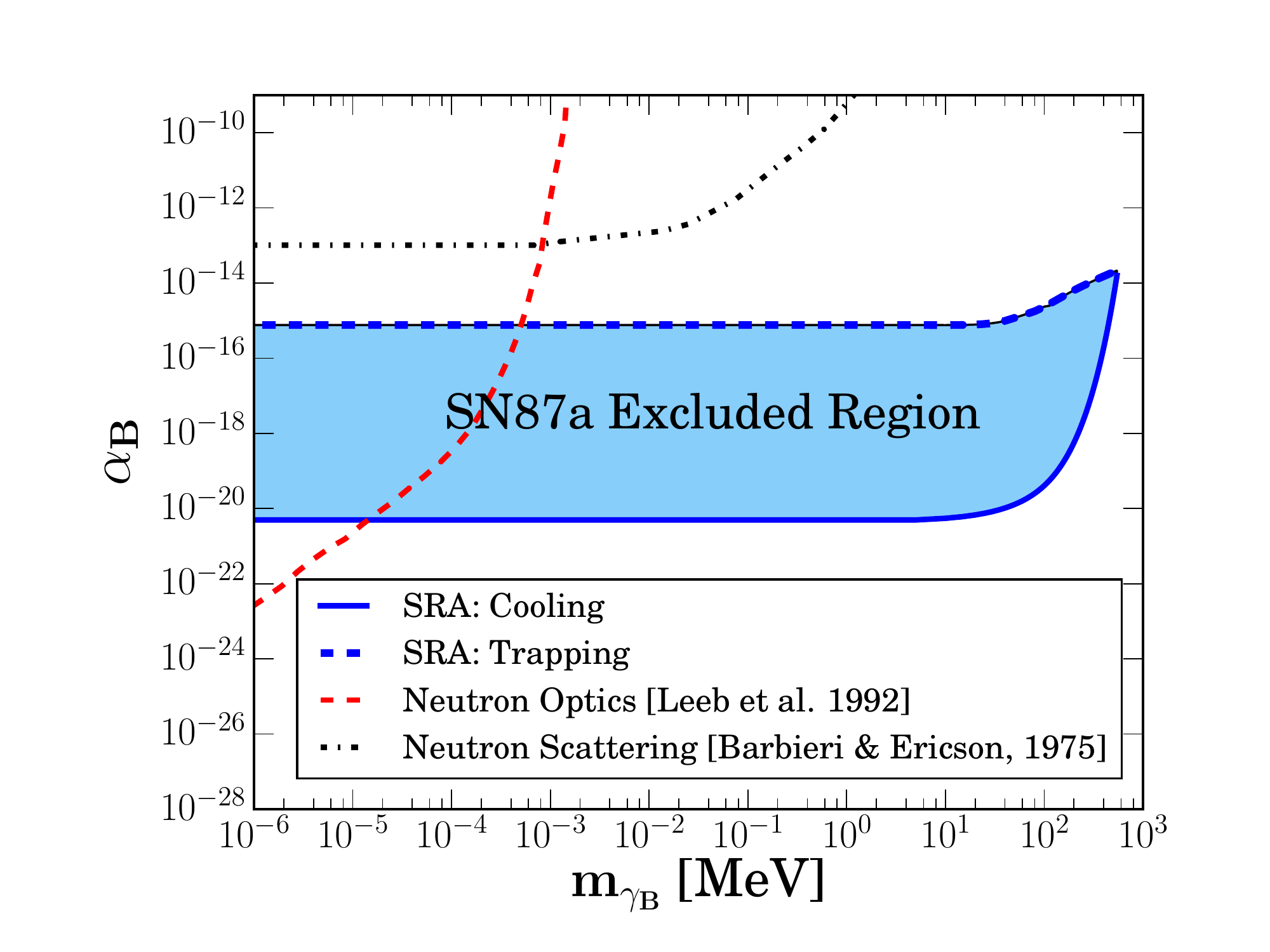}
 \caption{Cooling and trapping constraints in the parameter space of the LVB that couples to baryon number. The solid blue line is the lower limit set by cooling,  and the dashed blue line is the upper limit set by trapping. Experimental constraints derived from neutron scattering from Ref.~\cite{Barbieri:1975xy} (black dot-dashed curve) and from neutron optics from Ref.~\cite{Leeb:1992qf} (red dashed curve) are also shown.}
 \label{fig:alphaB}
\end{figure}
\end{center}

In Fig.~\ref{fig:alphaB} we show the constraint on the coupling strength defined as $\alpha_B=\epsilon^2_B \alpha_{\rm em}$ where   $\alpha_{\rm em}=1/137$ is the fine structure constant. We have opted to work with $\alpha_B$ rather $\epsilon_B$ because this is widely used in the context of discussing LVBs that couple to baryon number. The solid blue curve is obtained by setting $\dot{E}_B( \rho=3\times 10^{14} ~\text{g}/ \text{cm}^{3}, T=30 ~\text{MeV}, Y_p=0.3)= 10^{19}~ \text{erg}/\text{g}/\text{s}$ and solving for $\epsilon_B$ for a range of LVB masses $m_B=1~\text{eV}-200~\text{MeV}$. For value of $\alpha_B$ larger than those defined by the blue curve the supernova  would cool too rapidly to produce the neutrino events detected from SN87a. For lighter masses when $m_B \ll 1~\text{eV}$ the exchange of the LVB leads to macroscopic forces, collectively referred as fifth forces, and have been probed by a host experiments (for a review see Ref.~\cite{Adelberger:1991}). These have strongly constrained $\alpha_B$  to values that are several orders of magnitude smaller than can be accessed by the SN cooling constraint. At intermediate values in the range $m_B \simeq {\rm few}~\text{eV}-\text{MeV}$ neutron scattering and neutron optics provide the strongest experimental constraints \cite{Barbieri:1975xy,Leeb:1992qf} and these are also shown in Fig.~\ref{fig:alphaB}. 

While it is remarkable that the SN cooling constraint in Fig.~\ref{fig:alphaB} is several  orders of magnitude more stringent than the experimental constraints it relies on the assumption that once produced the LVBs can free stream out of the proto-neutron star.  Clearly this will not be true for large values of the coupling $\alpha_B$. At these larger values of $\alpha_B$ LVBs will be trapped in the core and will be emitted as black-body radiation with a luminosity 
\be
L_s \simeq \frac{\pi^3}{30} g^*(\eta)~R^2_s  T^4_s
\label{eq:L_s}
\ee
where $\eta=m/T_s$, $R_s$ and $T_s$ are the radius and temperature at which LVBs decouple and the effective spin degree of freedom 
\be 
g^*(\eta)=\frac{45}{\pi^4} \int^\infty_0 dy~\frac{y^2\sqrt{\eta^2+y^2}} {\exp{(\sqrt{\eta^2+y^2})}-1}
 \ee
is the correction to it which includes the effects due to the finite mass of LVBs and the contribution from the additional longitudinal modes.  When  $L_s > 3\times 10^{52}$ ergs/s  the neutrino signal  is discernibly altered and neutrino events from SN87a provide an upper bound on the coupling of LVBs \cite{Raffelt:1996wa}.   

An accurate description of the decoupling process will rely on computer simulations of proto-neutron star evolution which include energy transport due to both neutrinos and LVBs and is beyond the scope of this study. In what follows we shall adopt a simple criterion to estimate the upper bound on our constraint.  For LVB radiation from  the decoupling surface to discernibly reduce the neutrino luminosity, a large fraction of this radiation originating at $R_s$ should propagate to regions beyond the radii at which neutrinos decouple from matter. Otherwise, the energy radiated in LVBs will be reabsorbed by matter and transferred back to neutrinos. To enforce this we define an effective optical depth in the vicinity of $R_s$ 
\be 
\tau(R_s)=\int^{R_m}_{R_s} \frac{dr}{\langle {\lambda(r)} \rangle }\,,
\label{eq:decoupling} 
\ee 
where
\be 
\langle {\lambda(r)} \rangle = \frac{\int^\infty_\eta~dx ~\frac{x^2\sqrt{x^2-\eta^2} }{(e^{x}-1)}~\lambda_r(\omega=x T_s)}{\int^\infty_\eta~dx ~\frac{x^2\sqrt{x^2-\eta^2} }{(e^{x}-1)}}
\label{eq:mean_lambda}
\ee
is a simple energy weighted spectral average of the mean free path $\lambda_r(\omega)$ of LVBs, $\eta=m/T_s$ and $R_m$ is the radius at which the temperature has dropped to $T=T_\nu/2\approx 3$ MeV, and require that $\tau(R_s) < 3$. The choice of $R_m$ and $\tau(R_s)$ are well motivated but the associated errors are difficult to asses because they are compounded by the fact that the ambient conditions in the vicinity of $R_s$ change with time and in analysis here we use static profiles of density and temperature. Near the surface of the newly born neutron star density and temperature can be modeled using simple power laws given by  $\rho(r)=\rho(R_s) (R_s/r)^n$ and $T(r)=T_s(R_s/r)^{n/3}$.  The index $n$ is varied over the range 3--7 it is possible to mimic representative profiles found from supernova and proto-neutron star simulations at a characteristic time of 1--2 seconds after bounce from Ref.~\cite{Keil:1994sm,Pons:1998mm}.  However when the mass of the LVB is is much larger than the temperature in the outer regions, the decoupling surface will be pushed to higher temperature in the core. To describe this heavier mass decoupling  we smoothly connect the steep surface profiles at the surface to a slowly varying density in the core $\rho_{\rm core}\simeq 3 \times 10^{14}$ g/cm$^3$ and nearly constant core temperature $T_{\rm core}=30$ MeV.  

\begin{center}
 \begin{figure}[ht]
  \includegraphics[scale=0.5]{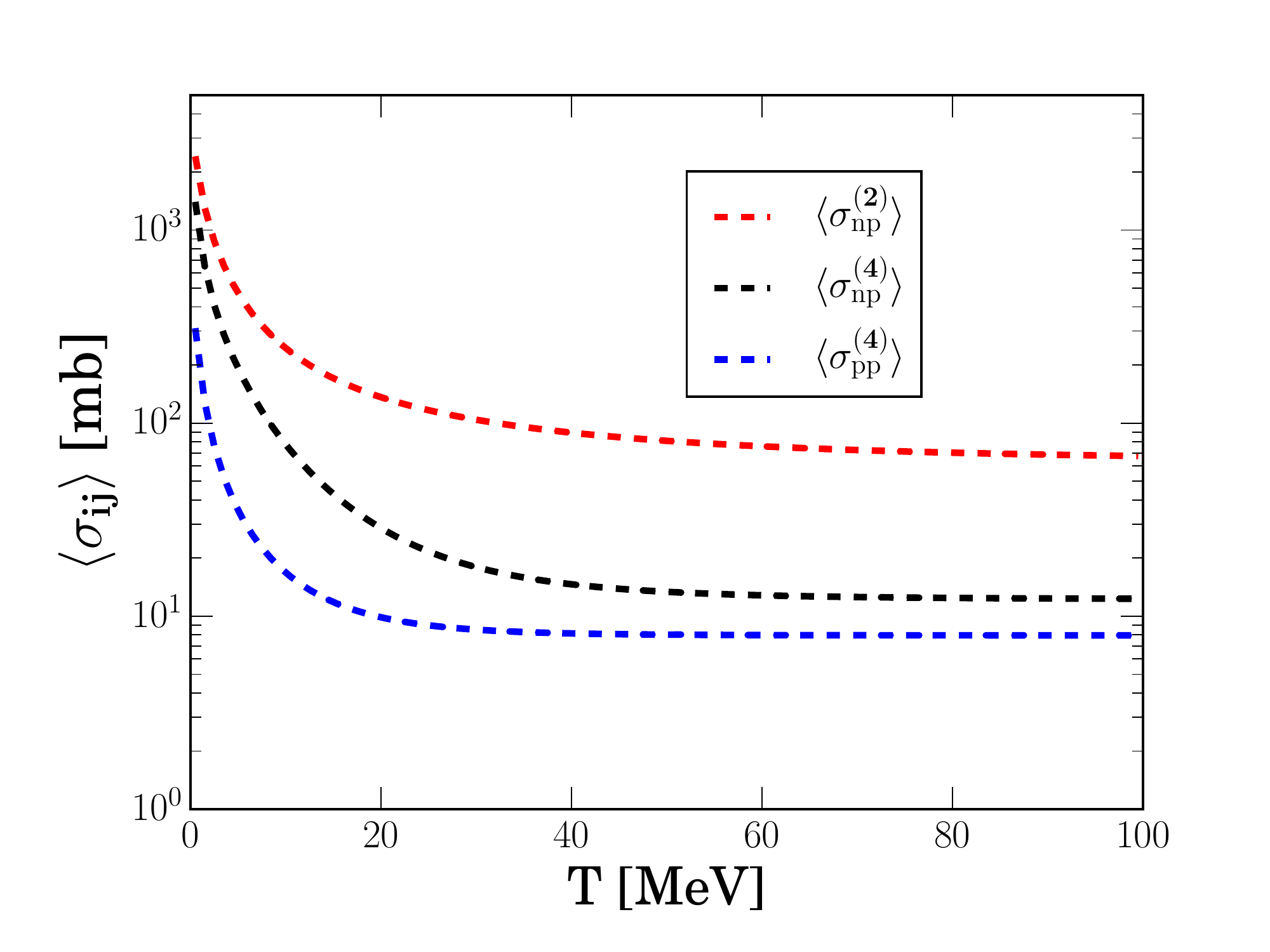}
  \caption{The temperature dependence of the thermally averaged nuclear cross sections needed for the calculation of the bremsstrahlung absorption contributions to the mean free path of LVBs.}
  \label{fig:sigmat}
 \end{figure}
\end{center}

For densities and temperatures of interest, inverse bremsstrahlung reactions $\gamma_B np\rightarrow np$, $\gamma_B nn\rightarrow nn$  and $\gamma_B pp\rightarrow pp$ are more important than the Compton scattering process $\gamma_B p\rightarrow p \gamma$ (interestingly, due to plasma effects, Compton scattering off electrons $\gamma_B e^-\rightarrow e^- \gamma$ and pair production of electron-positron pairs $\gamma_B \rightarrow e^+ e^-$ is induced through in-medium mixing with the photon due a proton-hole loop but was found to be small compared to the bremsstrahlung processes). The mean free path due to the inverse bremsstrahlung  process can be calculated in the soft radiation approximation. Using the transition matrix element calculated for bremsstrahlung and making appropriate changes to the phase space integrals (see Appendix \ref{appendix:mfp} for details) we find the mean free path for the process  $\gamma_B ij\rightarrow ij$
\be
  \frac{1}{\lambda^{ij}_{\gamma_B}(\omega)}=\frac{2496}{135\pi}~\alpha_B~n_in_j~\left(\frac{\pi T}{M}\right)^{5/2} \frac{1}{\omega^3}\sqrt{1-\xi^2}(1+\frac{2}{13} \xi^2) ~ \langle \sigma^{(4)}_{ij}(T) \rangle 
 \label{eq:lambda_B}
\ee
where $n_i$, $n_j$ and number densities of the nucleons involved, $T$ is the ambient temperature, $\xi=m_B/\omega$, and the thermal cross-section 
\bea
\langle \sigma^{(4)}_{ij}(T) \rangle &=&\frac{1}{6}\int_{0}^{\infty}dx~e^{-x} x^3 ~\sigma^{(4)}_{ij}(\Ecm=x T)\,,
\label{eq:sigma_4}  \\
\sigma^{(4)}_{ij}(\Ecm)&=&\int_{-1}^{1}d\cos\theta_{\text{cm}}~(1-\cos^2\theta_{\text{cm}})~\frac{d\sigma_{ij}(\Ecm)}{d\theta_{\text{cm}}}\,.
\eea
Here, as before $d\sigma_{ij}(\Ecm)$ is the differential cross-section for elastic nucleon-nucleon scattering process $ij\rightarrow ij$ and $\Ecm$ is the center of mass energy of the nucleon pair in the initial state. In Fig.~\ref{fig:sigmat} the variation of the thermal cross sections with temperature is shown and the large increase at low temperature arises because the nucleon-nucleon cross sections at low energy increase rapidly, as can be seen in Fig.~\ref{fig:sigma_np_compare}. 
We note that  the one pion exchange model for nuclear interactions would have predicted the opposite behavior.  The thermal cross section which will be relevant when we discuss the opacity of dark photons later    
\bea
\langle \sigma^{(2)}_{np}(T) \rangle &=&\frac{1}{2}\int_{0}^{\infty}dx~e^{-x} x^2 ~\sigma^{(2)}_{np}(\Ecm=x T)\,,
\label{eq:sigma_2}  \\
\sigma^{(2)}_{np}(\Ecm)&=&\int_{-1}^{1}d\cos\theta_{\text{cm}}~(1-\cos\theta_{\text{cm}})~\frac{d\sigma_{np}(\Ecm)}{d\theta_{\text{cm}}}\,.
\eea
is also shown in Fig.~\ref{fig:sigma_np_compare}. It is interesting to note that $\langle \sigma^{(4)}_{ij}(T) \rangle $  relevant for LVBs that couple to baryon number is quite smaller because in this case scattering occurs only due quadrupole fluctuations of baryon charge in nucleon-nucleon collisions.   
Summing over the individual contributions the mean free path in Eq.~\ref{eq:mean_lambda} is given by $\lambda_r(\omega)=(1/\lambda^{np}_r+1/\lambda^{nn}_r+1/\lambda^{pp}_r)^{-1}$ and we use it in Eq.~\ref{eq:decoupling} and employ the matter profile previously mentioned to calculate $\tau(R_s)$. The blue dashed curve in Fig.~\ref{fig:alphaB} is the obtained by solving the $\tau(R_s)=3$. 
\begin{center}
 \begin{figure}[h]
  \includegraphics[scale=0.6]{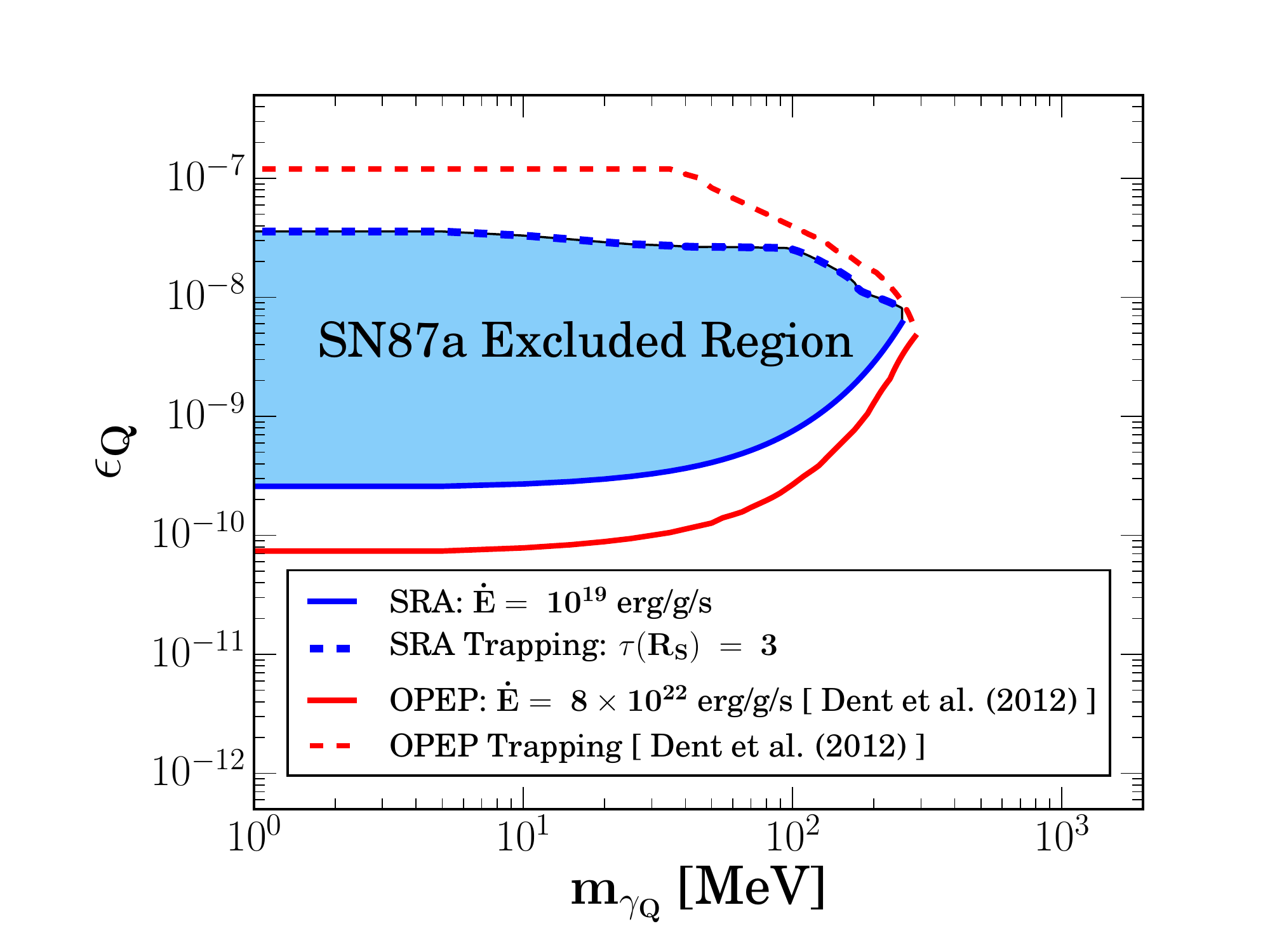}
  \caption{The revised excluded region in the dark photon parameter space. Blue curves show results obtained using the bremsstrahlung rates calculating in the SRA while the red curves are taken from \cite{Dent:2012mx} and are based on rates calculated using the OPEP (see text for details).}
  \label{fig:epsilon_Q}
  \end{figure}
\end{center}
We now turn to perform a similar analysis to constrain the properties of the dark photon. In this case the total energy loss rate per gram due to the radiation of dark photons is 
\be
\dot{E}_Q(\rho,T,Y_p) = (\dot{\epsilon}_{np\rightarrow np\gamma_Q}+\dot{\epsilon}_{pp\rightarrow pp\gamma_Q})/\rho \,,
\ee
and requiring that $\dot{E}_Q( \rho=3\times 10^{14} ~\text{g}/ \text{cm}^{3}, T=30 ~\text{MeV}, Y_p=0.3)< \dot{E}_{\rm Raffelt}$ provides a constraint on $\epsilon_Q$ that is shown by the solid blue curve in Fig.~\ref{fig:epsilon_Q}.  In this case, we choose to show constraints on  $\epsilon_Q$ rather than the related quantity $\alpha_Q= \epsilon^2_Q \alpha$ to help compare with earlier constraints obtained in Ref.~\cite{Dent:2012mx,Kazanas:2014mca}. For reference the SN cooling constraint from Ref.~\cite{Dent:2012mx} is also shown in Fig.~\ref{fig:epsilon_Q} as a solid red curve. The differences between the results arise due to two unrelated factors which partially offset each--other.
First, as noted earlier, the OPEP used in \cite{Dent:2012mx} to calculate the bremsstrahlung rate of dark photon production is expected to be larger than our predictions based on the SRA. In addition, the inclusion of a specific contribution to the the 2-body current coming from the pion-exchange current could spuriously enhance the $np$ bremsstrahlung rate by a large factor. 
Secondly, in \cite{Dent:2012mx} the authors chose to derive the constraint by requiring that total energy loss due to dark photons produced in the central $1$ km region of the SN core with a density $\rho_{\rm core}=3\times 10^{14} ~\text{g}/ \text{cm}^{3}$ and  $T_{\rm core}=30 ~\text{MeV}$ be less than $10^{53}$ ergs/s. This corresponds to a local bound on the energy loss $\dot{E} < 8\times 10^{22}$ ergs/g/s. Had we used this bound instead of $\dot{E}< \dot{E}_{\rm Raffelt}$ , our constraint on $\epsilon_Q$ would be weaker by a factor of about 100. This suggests that the bremsstrahlung rate in  \cite{Dent:2012mx} is larger than ours by a factor of about 500! This large difference cannot be explained by the differences we see between the data and the nucleon-nucleon cross sections predictions by the OPEP. It is also unlikely that inclusion of a specific meson-exchange contribution in \cite{Dent:2012mx} can account for this large enhancement.        

To obtain the upper bound on $\epsilon_Q$ due to trapping we calculate  the dark photon mean free path. At high density and for dark photon masses $m_Q < 100$ MeV the dominant absorption process is the inverse bremsstrahlung  $\gamma_Q np \rightarrow np$ and the associated mean free path is given by  
\be
  \frac{1}{\lambda^{np}_{\gamma_Q}(\omega)}=8\alpha_{\rm em}\epsilon_Q^2~n_in_j~\left(\frac{\pi T}{M}\right)^{3/2} \frac{1}{\omega^3}\frac{2+\xi^2}{\sqrt{1-\xi^2}}~ \langle \sigma^{(2)}_{np}(T) \rangle 
 \label{eq:lambda_Qnp}
\ee
where as before $n_n$, $n_p$ are number densities of neutrons and protons, $T$ is the ambient temperature, $\xi=m_Q/\omega$ and thermal cross-section was defined earlier in Eq.~\ref{eq:sigma_2}.  The reaction $\gamma_Q pp \rightarrow pp$ makes a smaller contribution because it occurs at quadrupole order and is given by 
\be
  \frac{1}{\lambda^{pp}_{\gamma_Q}(\omega)}=\frac{2496}{135\pi}\alpha_{\rm em}\epsilon_Q^2~n^2_p~\left(\frac{\pi T}{M}\right)^{5/2} \frac{1}{\omega^3}\sqrt{1-\xi^2}(1+\frac{2}{13} \xi^2)~ \langle \sigma^{(4)}_{pp}(T) \rangle \,,
 \label{eq:lambda_Qpp}
\ee
where the thermal cross section $ \langle \sigma^{(4)}_{pp}(T) \rangle$ was defined earlier in Eq.~\ref{eq:sigma_4}.  The direct decay of the dark photon to electron-positron pairs $\gamma_Q\rightarrow e^+ + e^-$ becomes relevant for larger dark photon masses. In the supernova core Pauli blocking of the final state electrons due to their high chemical potentials $\mu_e \simeq 100$ MeV suppresses this decay rate and the mean free path is given by 
\begin{equation}
 \begin{split}
 \frac{1}{\lambda_{\gamma_Q}^{e^{+}e^{-}}(\omega)}=&\frac{\alpha \epsilon^2}{3} \frac{m_Q^2+2m_e^2}{\omega^2-m_Q^2}\int_{E^{-}}^{E^{+}} d E~ (1-f_{e^-}(E))\\
  =&\frac{\alpha \epsilon^2}{3} \frac{m_Q^2+2m_e^2}{\omega^2-m_Q^2} \big[2(E^{+}-E^{-})-T\log(\frac{e^{E^{+}/T}-e^{\mu_e/T}}{e^{E^{-}/T}-e^{\mu_e/T}}) \big]
 \end{split}
\end{equation}
where,
\be
E^{\pm}= \Big[m_Q^2+\omega^2\Big(\sqrt{1-4\frac{m_e^2}{m_Q^2}}\pm \frac{1}{2}\sqrt{1-\frac{m^2_Q}{\omega^2}}\Big)^2\Big]^{1/2}\,,
\ee 
and $\omega=\sqrt{k^2 +m_Q^2}$ is the energy of the dark photon in the rest frame of the dense plasma. Including inverse bremsstrahlung and decay contributions the mean free path of the dark photon $\lambda_r(\omega)= (1/  \lambda_{\gamma_Q}^{e^{+}e^{-}}(\omega) + 1/\lambda^{pp}_{\gamma_Q}(\omega) + 1/\lambda^{np}_{\gamma_Q}(\omega))^{-1}$. We use this to calculate the optical depth defined in Eq.~\ref{eq:decoupling} and obtain the trapping upper bound on the constraint. As discussed earlier for larger values of $\epsilon_Q$  dark photons are reabsorbed in the region in the vicinity of the neutrino sphere and the neutrino emission will not be altered significantly.   

\section{Conclusions}
\label{sec:Conclusions}
We have calculated for the first time the energy loss rate due to dark gauge bosons that couple to baryon number from the supernova core and used it to  constrain its properties. We find that for gauge boson masses in the range $m_B=10^{-4}-10^{2}$ MeV the SN provides the most stringent constraint to date on the effective baryon number fine structure constant $\alpha_B$ which is about $6$ orders of magnitude smaller than earlier constraints based on neutron scattering data. Our calculation is based on the SRA which is valid in the limit when the energy carried by the radiation is small compared to the energy of nucleons involved in the reaction. In this limit the bremsstrahlung rate can be related to the nucleon-nucleon elastic scattering cross-sections and provides a benchmark that is independent of the potential used to model the nucleon-nucleon interaction. Using the SRA  we have also calculated the emissivity of dark photons and compared our predictions to those obtained in \cite{Dent:2012mx} which were based on the one pion exchange potential of nucleon-nucleon interactions. We find significant differences because one-pion exchange is a poor approximation to the nucleon-nucleon interaction. The revised SN cooling constraint for the dark photons is about one order of magnitude weaker.   

We have also calculated the LVB mean free paths in the SRA to estimate the upper bound on the coupling. We find that inverse bremsstrahlung reactions dominate the opacity. At the relevant densities and temperatures the SRA predicts an enhancement of these rates in the outer regions of the supernova when compared to the results obtained in the one pion exchange model because the latter underestimates the nucleon-nucleon elastic cross-section at low energy. 

While the SRA is a significant improvement over simple models of the nucleon-nucleon interaction treated in the Born approximation, its strictly valid for low energy processes where the expansion parameter $\chi=\omega /\Ecm \ll 1$. In our calculations we used the SRA for values of $\chi \simeq 1/5 - 1$ and the contribution of higher order terms in this expansions cannot be ignored. Nonetheless, as we noted earlier comparisons between the predictions of SRA and experimental data in the context of photon bremsstrahlung from nucleon-nucleon collisions have shown that the agreement between SRA predictions and data for photon energies $\lesssim 100$ MeV is typically better than expected, differing by about a factor of 2 at the higher energies. For these larger energies two-body currents and re-scattering diagrams contribute and their inclusion relies on a model of the nucleon-nucleon interaction and the associated 2-body currents. Chiral nucleon-nucleon potentials inspired by effective field theory are well suited for this purpose, and it would be desirable to first perform bremsstrahlung calculations in this framework, benchmark them with available data from $pp$ and $np$ bremsstrahlung experiments, and then employ them to predict the emissivities of LVBs in the supernova context. In addition, other corrections of ${\cal O}(1)$ arising from many-body effects in the dense core also need to be studied. If these corrections can suppress the emissivities in the core then the ability of supernovae to constrain LVBs will be further diminished. 

Further work is warranted before we can draw firm conclusions about the extent of the LVB parameter space excluded by SN87a.  The rapid increase in  nucleon-nucleon cross-sections at low energy implies that the  opacity due to the inverse bremsstrahlung processes in the outer cooler regions of the star is larger, while in contrast the smaller cross section at high energy imply a smaller emissivity in the high temperature core. Together, this indicates that the inclusion of realistic nuclear physics acts to reduce the region of parameter space that can be constrained. This underscores the need to further improve the nuclear physics of bremsstrahlung processes by going beyond the SRA, and to incorporate them self-consistently into supernova simulations. Corrections to the SRA would be especially relevant for LVB masses greater than about $100$ MeV so our constraints in the region of parameter space must be viewed as preliminary. Interestingly, for the dark photon masses in the MeV - GeV range and for coupling $\epsilon_Q$ in the range $10^{-7}-10^{-9}$, recent work suggests dark mater annihilations in the earth's core can lead to detectable signatures in terrestrial detectors \cite{Feng:2015hja}. Since this has significant overlap with the region constrained by SN87a it would be worthwhile to refine these constraints. As a first step we are including the contribution of LVBs in the energy transport of 1-d models of core-collapse supernova simulations using the formulae for the emissivities and opacities derived in this study and will be reported in future work.       
\begin{acknowledgments}
We would like to thank Michael Graesser, David Kaplan, Ann Nelson, Maxim Pospelov, Rob Timmermans and Martin Savage for useful discussions. The work of E. R. and S. R.was supported by the DOE Grant No. DE-FG02-00ER41132 and by the NUCLEI SciDAC program.
\end{acknowledgments}
\appendix
\section{Emissivity}
\label{appendix:emissivity}
The non-relativsitc limit of the current products ($J_{\mu}J^{\mu},\ L_{\mu}L^{\mu}$) is quadratic and quartic respectively in baryon velocity. 
The difference is due to an anomaly in the scenario where all baryon can radiate, as the center of charge is also the center of mass
and the dipole radiation vanishes and the leading term is quadrupole radiation. Another consequence of this difference is the dependence
on $\theta_{\text{cm}}$ as we show here.

Due to the boson being massive, the 4-vector product differs from the usual expression for photons:
\begin{equation}
 \begin{split}
  (\epsilon_{\mu}\tilde{J}^{\mu})^2 =&-(g_{\mu\nu}-\frac{k_{\mu}k_{\nu}}{m_A^2})\tilde{J}^{\mu}\tilde{J}^{\nu}\\
  =&-\tilde{J}_{\mu}\tilde{J}^{\mu}+(k_{\mu}\tilde{J}^{\mu})^2/m^2
 \end{split}
\end{equation}
where $m$ is the mass of LVB and $M = 938.918$ MeV is the average baryon mass. 

The term prortional to the currents:
\begin{equation}
 \begin{split}
  L_{\mu}L^{\mu}=&-\big[\frac{M^2}{(\bold{p}_1\cdot\bold{k})^2}+\frac{M^2}{(\bold{p}_3\cdot\bold{k})^2}-2\frac{(\bold{p}_1\cdot\bold{p}_3)}{(\bold{p}_1\cdot\bold{k})(\bold{p}_3\cdot\bold{k})}\big]\\
 J_{\mu}J^{\mu}=&-\big[\sum_{i=1}^4\frac{M^2}{(\bold{p}_i\cdot\bold{k})^2}+\frac{(\bold{p}_1\cdot\bold{p}_2)}{(\bold{p}_1\cdot\bold{k})(\bold{p}_2\cdot\bold{k})}+\frac{(\bold{p}_3\cdot\bold{p}_4)}{(\bold{p}_3\cdot\bold{k})(\bold{p}_4\cdot\bold{k})}-\sum_{i=1}^2\sum_{j=3}^4\frac{(\bold{p}_i\cdot\bold{p}_j)}{(\bold{p}_i\cdot\bold{k})(\bold{p}_j\cdot\bold{k})}\big]\\
 \label{eq:JLcurrents}
 \end{split}
\end{equation}
Due to SRA, the scattering is almost elastic, and the mass dependent term does not contribute:
\begin{equation}
 \begin{split}
  J_{\mu}^{np}k^{\mu}=&\big[\frac{\bold{p}_1\cdot\bold{k} }{\bold{p}_1\cdot\bold{k}}-\frac{\bold{p}_3\cdot\bold{k} }{\bold{p}_3\cdot\bold{k}}\big]=0\\
  J_{\mu}^{np}k^{\mu}=&\big[\frac{\bold{p}_1\cdot\bold{k} }{\bold{p}_1\cdot\bold{k}}+\frac{\bold{p}_2\cdot\bold{k} }{\bold{p}_2\cdot\bold{k}}-\frac{\bold{p}_3\cdot\bold{k} }{\bold{p}_3\cdot\bold{k}}-\frac{\bold{p}_4\cdot\bold{k} }{\bold{p}_4\cdot\bold{k}}\big]=0\\
 \end{split}
\end{equation}
In the non-relativsitc limit the velocities are defined as $\vec{v}_i=\vec{p}_i/M$ and energies are $E_i = M \sqrt{1+v_i^2}\approx M(1+v_i^2/2)$.
The expansion of Eq.~\ref{eq:JLcurrents} to leading order in velocities depends on the relative momentum of the incoming and outgoing nucleons:
\begin{equation}
 \begin{split}
 L_{\mu}L^{\mu}=\frac{1}{\omega^2}\Big\{&\frac{(\vec{p}-\vec{p}\ {'})^2}{M^2}-\big[\frac{(\vec{p}-\vec{p}\ {'})}{M}\cdot \frac{\vec{k}}{\omega}\big]^2\Big\}\\
  J_{\mu}J^{\mu}=\frac{4}{\omega^2}\Big\{&(\frac{\vec{p}\ {'}}{M}\cdot \frac{\vec{k}}{\omega})^2[\frac{p{'}^2}{M^2}+2 (\frac{\vec{p}}{M}\cdot \frac{\vec{k}}{\omega})^2]+\frac{p^2}{M^2}(\frac{\vec{p}}{M}\cdot \frac{\vec{k}}{\omega})^2\\
  &-\big[\frac{(\vec{p}+\vec{p}\ {'})}{M}\cdot \frac{\vec{k}}{\omega}\big]^4 -2\frac{(\vec{p}\cdot \vec{p}\ {'})}{M^2}(\frac{\vec{p}}{M}\cdot \frac{\vec{k}}{\omega})(\frac{\vec{p}\ {'}}{M}\cdot \frac{\vec{k}}{\omega})\Big\}\\
  \label{eq:vexpansion}
 \end{split}
\end{equation}
where, $\vec{p}=\frac{1}{2}(\vec{p}_1-\vec{p}_2),\ \vec{p}\ {'}=\frac{1}{2}(\vec{p}_3-\vec{p}_4)$, are the usual relative momenta in the center of mass frame.

In the soft limit $p = p\ {}'$. Performing the integration of the relative angles between the emitted LVB and the nucleons  confirms
the dipole and quadrupole radiation statement made earlier as can be seen in Eq.~\ref{eq:angles}.
\begin{equation}
 \begin{split}
  \frac{\omega^2}{4\pi}\int d\Omega_{\omega}L_{\mu}L^{\mu}=& 2\frac{E_{\text{CM}}}{M}(1-\frac{k^2}{3\omega^2})  (1 -\cos \theta_{\text{CM}})\\
  \frac{\omega^2}{4\pi}\int d\Omega_{\omega}J_{\mu}J^{\mu}=&\frac{8}{15} (\frac{E_{\text{CM}}}{M})^2 \frac{k^2}{3\omega^2} (5 - 2 \frac{k^2}{3\omega^2})  (1-\cos^2 \theta_{\text{CM}})\\
 \label{eq:angles}
 \end{split}
\end{equation}
Since $n-p$ cross section is coupled to dipole radiation,
its contribution is dominat in determining the rate of emission. 
This rather simple feature has important consequences, as the supression of emission from quadrupole radiation provides weaker cosntraints on LVBs 
coupled to baryon number with respect to the same ambient conditions for LVB coupled to electric charge. The final expressions in Eq.~\ref{eq:currents} are 
obtained by integrating out of Eq.~\ref{eq:angles} the radiated energy. 
\section{Unpolarized differential cross-section}
\label{section:appendix}
The scattering amplitude in the helicity basis, $M^{s',s}_{m_s',m_s}$,  is described in detail in \cite{PWA}.
Here, we relate it to the unpolarized differential cross section needed for our calculations:
\begin{equation}
 \begin{split}
  \frac{d \sigma^{el,un}}{d\Omega} =&\frac{1}{4} \text{Tr}M M^{\dagger}\\
 \end{split}
\end{equation}
From time reversal invariance:
\begin{equation}
 \begin{split}
  M^{1,1}_{1,1}= M^{1,1}_{-1,-1},\ M^{1,1}_{1,1}= M^{1,-1}_{-1,1},\ M^{1,1}_{0,1}= -M^{1,1}_{-1,0},\ M^{1,1}_{1,0}= -M^{1,1}_{0,-1} 
 \end{split}
\end{equation}
Thus,
\begin{equation}
 \begin{split}
  \frac{d \sigma}{d\Omega} =& \frac{1}{4}\big\{2|M^{1,1}_{1,1}|^2+|M^{1,1}_{0,0}|^2+|M^{0,0}_{0,0}|^2+2|M^{1,1}_{1,0}|^2+2|M^{1,1}_{0,1}|^2+2|M^{1,1}_{1,-1}|^2\big\}
 \end{split}
\end{equation}
Each matrix element can be expanded in partial wave basis as follows,
{\fontsize{10}{1}\selectfont
\begin{equation}
 \begin{split}
 M^{0,0}_{0,0}=&(ip)^{-1}\sum_LP_L (\frac{L+1}{2}) \alpha_{L,L}\\
  M^{1,1}_{0,0}=&(ip)^{-1}\sum_LP_L\Big[ (\frac{L+1}{2})\alpha_{L,L+1}+ (\frac{L}{2})\alpha_{L,L-1}+\frac{\sqrt{(L+1)(l+2)}}{2}\alpha^{l+1}+\frac{\sqrt{L(L-1)}}{2}\alpha^{L-1}\Big] \\
 M^{1,1}_{0,1}=& -(i p)^{-1}e^{i\phi}\sum_{L}P^1_L\Big [ \frac{\sqrt{2}}{4}(\frac{2L+1}{L(L+1)})\alpha_{L,L}+\frac{\sqrt{2}}{4}(\frac{L-1}{L})\alpha_{L,L-1}-\frac{\sqrt{2}}{4}(\frac{L+2}{L+1})\alpha_{L,L+1}+\frac{\sqrt{2}}{4}\sqrt{\frac{L+2}{L+1}}\alpha^{L+1}\\
 M^{1,1}_{1,0}=& -(i p)^{-1}e^{-i\phi}\sum_{L}P^1_L \Big [ \frac{\sqrt{2}}{4}\alpha_{L,L+1}-\frac{\sqrt{2}}{4}\alpha_{L,L-1}+\frac{\sqrt{2}}{4}\sqrt{\frac{L+2}{L+1}}\alpha^{L+1}-\frac{\sqrt{2}}{4}\alpha^{L-1}\Big]\\
  M^{1,1}_{1,-1}=& (i p)^{-1}e^{-2i\phi}\sum_{L}P^2_L \Big [\frac{\alpha_{L,L+1}}{4(L+1)}-\frac{2L+1}{4L(L+1)}\alpha_{L,L}+\frac{\alpha_{L,L-1}}{4L}-\frac{\alpha^{L+1}}{4\sqrt{(L+1)(L+2)}}-\frac{\alpha^{L-1}}{4\sqrt{L(L+1)}}\Big]\\
  \end{split}
\end{equation}}
The dependence on the nuclear-bar phase shifts ($\delta^{S,L,J}$) is given below:
\begin{equation}
 \begin{split}
  \alpha_{J,J}=&e^{2i\delta^{1,J,J}}-1\\
  \alpha_{J\pm1,J}=&\cos(2\epsilon_J)e^{2i\delta^{1,\pm J,J}}-1\\
  \alpha^J=&i \sin(2\epsilon_J) e^{i(\delta^{1,J+1,J}+\delta^{1,J-1,J})}\\
  \alpha_J=&e^{2i \delta^{0,J,J}}-1
 \end{split}
\end{equation}
For the p-p channel the total matrix elements have to be antisymmetrized. This means that the spatial component has to be symmetrized for spin singlet
and anti-symmetrized for spin triplet.
When performing angular integration, the following integral needs to be evaluated analytically:
\begin{equation}
 \begin{split}
  I_{l_1,\ l_2,\ l_3}^{m_1,m_2,m_3}\equiv&\frac{1}{2}\int_{-1}^{1}P_{l_1}^{m_1}(x)P_{l_2}^{m_2}(x)P_{l_3}^{m_3}(x)dx
 \end{split}
\end{equation}
In our specific case we need expression for $I^{m\ m\ 0}_{l_1\ l_2\ l_3},\ l_3=\{0,1,2\}, (l_1+l_2)\mid 2$, and we provide
the identities needed.
\begin{equation}
 \begin{split}
  I^{m\ m\ 0}_{l_1\ l_2\ 0}=& \frac{(l_1+m)!}{(2l+1)(l_1-m)!}\delta_{l_1, l_2}\\
  I^{m\ m\ 0}_{l_1\ l_2\ 1}\Big|_{|l_1-l_2|\leq 1}=&\frac{(-1)^{(l_1-l_2+1)/2}(l_2+m)!(l_1+l_2-1)!((l_1+l_2+1)/2)!}{(l_2-m)!((l_2-l_1+1)/2)!((l_1-l_2+1)/2)!((l_2+l_1-1)/2)!(l_2+l_1+2)!}\\
  &\times \sum_{t=\text{Max}(0,1-m-l_2)}^{\text{Min}(1+l_2-m,l_1-m,1)}\frac{(-1)^t(l_1+t+m)!(l_2+1-m-t)!}{t!(l_1-m-t)!(l_2-1+m+t)!(1-t)!}\\
  I^{m\ m\ 0}_{l_1\ l_2\ 1}\Big|_{|l_1-l_2|> 1}=&0\\
  I^{m\ m\ 0}_{l_1\ l_2\ 2}\Big|_{|l_1-l_2|\leq 2}=&\frac{2(-1)^{(l_1-l_2)/2+1}(l_2+m)!(l_1+l_2-2)!((l_1+l_2)/2+1)!}{(l_2-m)!((l_2-l_1)/2+1)!((l_1-l_2)/2+1)!((l_2+l_1)/2-1)!(l_2+l_1+3)!}\\
  &\times \sum_{t=\text{Max}(0,2-m-l_2)}^{\text{Min}(2+l_2-m,l_1-m,2)}\frac{(-1)^t(l_1+t+m)!(l_2+2-m-t)!}{t!(l_1-m-t)!(l_2-2+m+t)!(2-t)!}\\
  I^{m\ m\ 0}_{l_1\ l_2\ 2}\Big|_{|l_1-l_2|> 2}=&0
 \end{split}
\end{equation}
\section{Mean Free Paths}
\label{appendix:mfp}
The derivation of the mean free path is similar to the emissivity, with two main differences; there is no integration of the radiated energy,
and the focus is on particle absoprtion and not energy loss (there is a factor of energy missiing in comparison with Eq.~\ref{eq:emis_Q}, \ref{eq:emis_B}).
In equations~\ref{eq:gamma_Q} and \ref{eq:gamma_B} we show the deacy rate ($\Gamma = v_{\tilde{\gamma}}/\lambda$):
\bea 
{\Gamma}_{np\rightarrow np \gamma_Q} &=& -\frac{2 \pi}{\omega} \alpha_{\rm em} \epsilon_Q^2   \int \frac{d^3p_1}{(2E_1)(2\pi)^3} f_n(E_1) \int \frac{d^3p_2}{(2E_2)(2\pi)^3} f_p(E_2) \int d\Pi (\epsilon^{\mu} J^{(2)}_\mu)^2 ~32 \pi E^2_{\rm cm}~v_{rel}~\frac{d\sigma_{\rm np}(E_{\rm cm},\theta)}{d\theta_{\rm cm}} \,,
\label{eq:gamma_Q}
\\
{\Gamma}_{np\rightarrow np \gamma_B} &=& -\frac{2\pi}{\omega} \alpha_{\rm em} \epsilon_B^2  \int \frac{d^3p_1}{(2E_1)(2\pi)^3} f_n(E_1) \int \frac{d^3p_2}{(2E_2)(2\pi)^3} f_p(E_2) \int d\Pi (\epsilon^{\mu} J^{(4)}_\mu)^2 ~32 \pi E^2_{\rm cm}~v_{rel}~\frac{d\sigma_{\rm np}(E_{\rm cm},\theta)}{d\theta_{\rm cm}} \,.
\label{eq:gamma_B}
\eea
The non-relativistic expansion of the currents to leading order has been performed in Appendix~\ref{appendix:emissivity},
and after integrating over the angle between the center of mass momentum of one of the incoming nucleons and the 
momentum of the outgoing LVB we find 
\begin{equation}
\begin{split}
 \frac{1}{4\pi}\int d\Omega~L_{\mu}L^{\mu}\equiv&\frac{\Ecm}{M}\frac{\mathcal{J}^{(2)}}{\omega^2}(1-\cos\theta_{\rm cm})\\
 \frac{1}{4\pi}\int d\Omega~J_{\mu}J^{\mu}\equiv&\frac{\Ecm}{M}\frac{\mathcal{J}^{(4)}}{\omega^2}(1-\cos^2\theta_{\rm cm})\\
\end{split}
\end{equation}
where
\begin{equation}
 \begin{split}
  \mathcal{J}^{(2)}=&2(1-\frac{k^2}{3\omega^2})\\
  \mathcal{J}^{(4)}=&\frac{8}{45}~\frac{E_{\text{CM}}}{M}~\frac{k^2}{\omega^2}(5 -\frac{2k^2}{3\omega^2})\\
 \end{split}
\end{equation}
The integration over initial states of the nucleons can be simplified when they are non-degenerate and we find that 
\begin{equation}
 \int \frac{d^3p_1}{(2\pi)^3} \int \frac{d^3p_2}{(2\pi)^3} ~ f_i(E_1)f_j(E_2)=\frac{n_in_j}{2 \sqrt{\pi}~T^{3/2}}\int d\Ecm~ e^{-\Ecm/T}\sqrt{\Ecm}  \,. 
\end{equation}
Putting all these expressions back into Eq.~\ref{eq:gamma_Q} and Eq.~\ref{eq:gamma_B} the expressions for the mean free paths in our work can be easily derived.


\end{document}